\documentclass[12pt,preprint]{aastex}

\newcommand\msunyr{\rm M_{\odot}\,yr^{-1}}
\newcommand\msun{\rm M_{\odot}}

\newcommand\rsun{\rm R_{\odot}}
\newcommand\mdot{ \dot{M}}

\newcommand\be {\begin{equation}}
\newcommand\en{\end{equation}}
\newcommand\cm{\rm cm}

\def\'#1{\ifx#1i{\accent"13\i}\else{\accent"13#1}\fi}

\def\10alamenos#1{10$^{-#1}$}
\def\10ala#1{10$^{#1}$}

\def\cm2g{\rm cm^2 \ g^{-1}}

\begin{document}

\title
{The Effects of UV Continuum and Lyman $\alpha$ Radiation on the Chemical Equilibrium
of T Tauri Disks}

\author{Edwin Bergin \altaffilmark{1}, Nuria Calvet \altaffilmark{1},
Paola D'Alessio \altaffilmark{2}, and Gregory J. Herczeg\altaffilmark{3}}
\altaffiltext{1}{Harvard-Smithsonian Center for Astrophysics, 60 Garden St.,
Cambridge, MA 02138, USA;
Electronic mail: ebergin@cfa.harvard.edu, ncalvet@cfa.harvard.edu}
\altaffiltext{2}{Instituto de Astronom\'{\i}a,
UNAM, Apartado Postal 72-3 (Xangari), 58089 Morelia, Michoacan, Mexico;
M\'exico; Electronic mail:p.dalessio@astrosmo.unam.mx}
\altaffiltext{3}{
JILA, University of Colorado
Boulder, CO 80309-0440; Electronic mail: gregoryh@casa.colorado.edu}

\begin{abstract}
We show in this {\it Letter} that the spectral details of the FUV radiation 
fields have a large impact on the chemistry of protoplanetary disks 
surrounding T Tauri stars. We show that the strength of a realistic stellar 
FUV field is significantly lower than typically assumed in chemical 
calculations and that the radiation field is dominated by strong line 
emission, most notably Lyman $\alpha$ radiation. The effects of the strong 
Lyman $\alpha$ emission on the chemical equilibrium in protoplanetary disks 
has previously been unrecognized. We discuss the impact of this radiation
on molecular observations in the context of a radiative transfer model that 
includes both direct attenuation and scattering. In particular, Lyman 
$\alpha$ 
radiation will directly dissociate water vapor and may contribute to the 
observed enhancements of CN/HCN in disks.
\end{abstract}


\section{Introduction}
\label{sec_intro}

Over the past few years it has become evident that the 
protoplanetary disks surrounding newly formed stars contain a rich evolving chemistry,
which is a vital ingredient in the
overall structure and dynamics of the disk.  Because the dominant molecule (H$_2$) is
unexcited at the cold temperatures (T $\sim$ 10 -- 70 K) 
characteristic of most of the disk mass, trace species provide
cooling for upper layers that are unable to radiate via H$_2$ 
emission.  Moreover, the most promising process for the required angular momentum
transport is via the magnetorotational instability (Balbus \& Hawley 1991). This process
is critically tied to the abundance of charged particles that is set by the
penetration of energetic radiation and chemical reactions (Gammie 1996, Stone et al.
2000). 

At present,  emission from CO
has been detected in many systems \citep{dutrey_codisk, koerner_codisk},
while emission from other less abundant and more complex molecular species
has only been detected in a handful of sources
\citep{kastner_diskchem, dutrey_diskchem, goldsmith_ch3oh, qi_diskchem}.
Analysis of the observations reveals that the molecular
emission is arising primarily from the disk surface and that 
chemistry controlled by energetic ultraviolet (UV) and X-ray
radiation (which dissociates molecules and is active on surfaces) 
must contribute to the observed abundances \citep{aikawa_zchem1, willacy_zchem, vanz_submm}. 
 
Chemical models of these systems have been developed to 
examine the impact of stellar and interstellar (ISRF) FUV radiation on the
chemistry at the disk surface \citep{aikawa_zchem1, willacy_zchem,
aikawa_zchem2, vanz_zchem}.
These models typically assume that the stellar FUV field has the
same wavelength dependence as the ISRF,  and it is
$\sim 10^4$ stronger than the ISRF at 100 AU.
However, in this paper we argue that the strength of the FUV radiation field 
from a typical accreting T Tauri star (TTS) is much weaker than previously 
assumed and show that 
the observed stellar spectra do not have the same wavelength dependence as the ISRF. 

In addition, the presence of strong line radiation, especially Lyman~$\alpha$, has yet to 
be recognized in chemical models and may have important effects. 
For instance, nearly all T Tauri disks detected so far appear to have large ratios of 
CN relative to HCN (Dutrey et al. 1997; Kastner et al. 1997; Qi 2001).
These two species are linked to Lyman~$\alpha$ radiation 
because the main dissociative photoabsorption channels for CN are shortward 
of 1150 {\AA} (Nee \& Lee 1985); in contrast, HCN will be subject 
to photodissociation by Lyman~$\alpha$ photons near 1216 \AA\ (Nuth \& Glicker 1982).
The high CN/HCN ratio has been attributed to the photodissociation
by continuum FUV radiation from the star or the interstellar medium
(Aikawa et al. 1999; Willacy \& Langer 2000; van Zadelhoff et al. 2002).
However, strong Lyman~$\alpha$ fluxes may also contribute to elevate the CN/HCN ratio
and to the preferential destruction of other species including water vapor.

In this Letter we use FUV data of TTS to construct a 
FUV radiation field which is representative of low mass T Tauri stars
with average accretion properties. 
Using this field, we examine its effect on the
chemical equilibrium on the proto-planetary disk surface.

\section{The FUV spectrum of Classical T Tauri stars}
\label{sec_fuv}

In most calculations of chemical equilibrium in TTS disk
to date  it has been assumed that
the stellar FUV flux is $10^4$ times higher than the
ISRF at R = 100 AU (Willacy \& Langer 2000;
Aikawa et al. 2001). This value is taken from Herbig
\& Goodrich (1986) IUE observations of T Tau and RY Tau.
The FUV spectra below $\sim$ 2000 {\AA}
of these two stars is similar with 
$F_\lambda \sim 3 \times 10^{-13} {\rm erg \, cm^{-2} \, s^{-1}} $ {\AA}$^{-1}$,
yielding a flux at $R$ = 100 AU integrated over 
1000 {\AA} of $23.5 \, {\rm erg \, cm^{-2} \, s^{-1} } \sim 1.5 \times 10^{4}$
Habing (1 Habing $= 1.6 \times 10^{-3} {\rm erg \, cm^{-2} \, s^{-1}}$; 
Tielens \& Hollenbach 1985).  
However, there
are reasons to expect that these spectra do {\it not} represent
the typical FUV flux of TTS. For one thing,
T Tau and RY Tau
have larger masses and radii than 
typical TTS, and thus are not representative of the class \citep{KH95}.
In addition, the mass accretion
rate in their disks is $\mdot \sim 10^{-7} \msunyr$ 
(Calvet et al. 2003, in preparation),
higher than the mean TTS mass accretion
rate, $\sim 10^{-8} \msunyr$ \citep {HCGD}.
This is important, because the luminosity in the FUV
scales with accretion luminosity (Calvet et al. 2003).

\begin{figure}
\plotone{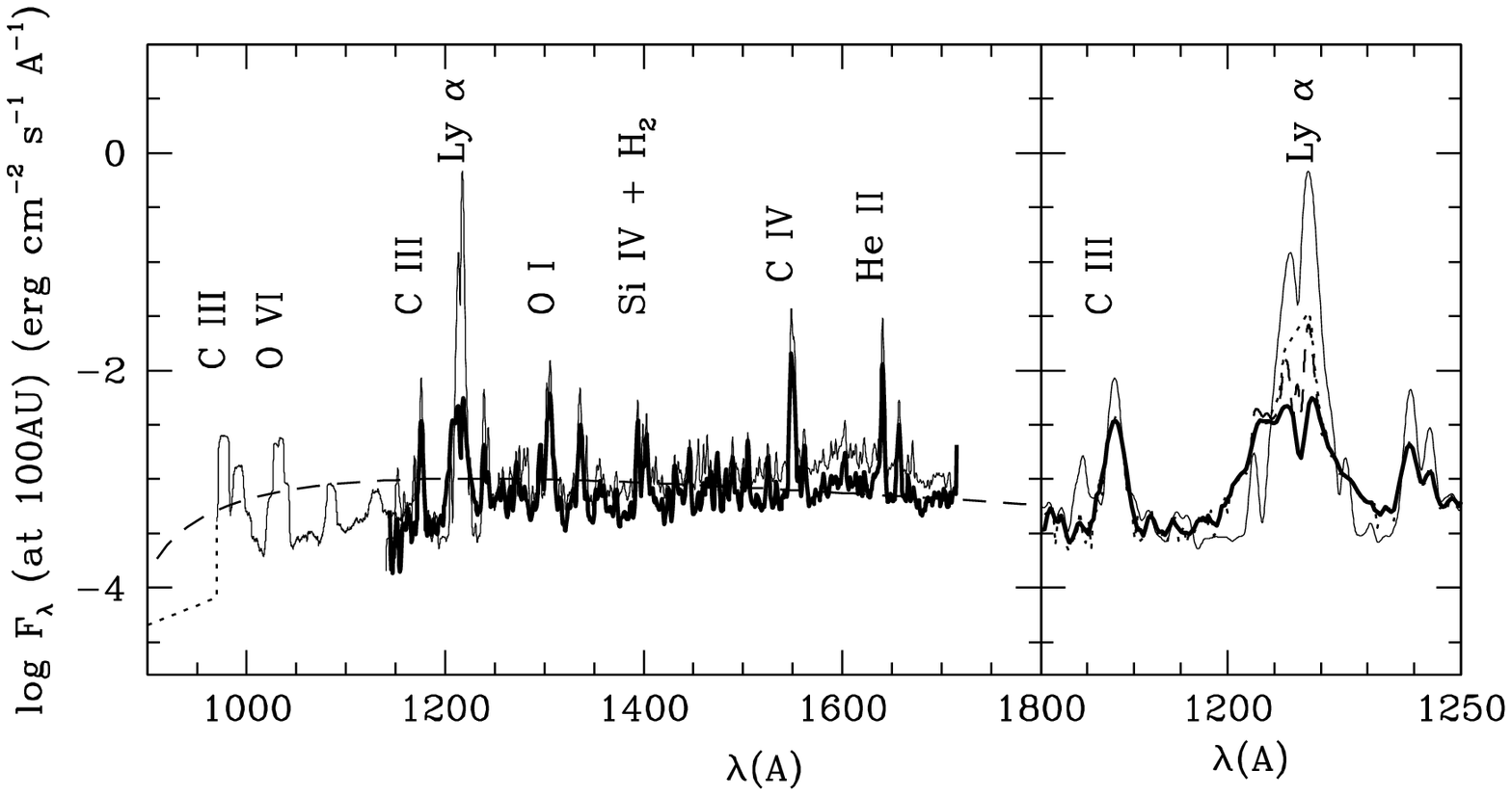}
\vskip -1in
\caption{Left: FUV spectra of T Tauri stars. Heavy solid line:
HST/STIS spectrum of BP Tau, obtained in program GO9081,
with grating G140L. Light solid line: spectrum of TW Hya scaled
by 3.5 to match the BP Tau continuum level. We have
combined the HST/STIS E140M spectrum from program GO8041, above $\lambda =$ 1150 {\AA},
with the FUSE spectrum below (Herczeg et al. 2002). The STIS
E140 M has been convolved to the G140L resolution, while the
FUSE spectrum has been smoothed by 10 {\AA}.
Identification of important emission lines
is given. The extension below 960 \AA\ adopted in the calculations for this work is 
shown as a dotted line (see \S \ref{sec_fuv}). The interstellar field from
Draine (1978) scaled
by $\sim$540 Habing to match the integrated flux of the
adopted spectrum is shown in dashed lines.
Right: region around the Lyman $\alpha$ line. The echelle
spectrum of CY Tau obtained with STIS and grating E140M in our program
GO8206 (Saucedo et al. 2003), convolved
to the resolution of G140L 
is shown with dashed lines. 
Other line styles are as in left panel.
The reddening towards CY Tau is lower than
towards BP Tau, and thus the central core of the line is more apparent.
The dotted line shows one of the adopted approximations for the
flux in the central core of the line (see \S \ref{sec_chemical}).
}
\label{tts}
\end{figure}

We now have a much better knowledge of the FUV fields
characteristic to TTS. 
Figure \ref{tts} shows the FUV spectrum
of BP Tau, a TTS with mass, radius, $\mdot$, and age  (0.5$\msun$,
2 $\rsun$, 
$ 3 \times 10^{-8} \msunyr$, and $\sim$ 2 Myr;  Gullbring et al. 1998)
typical of the class.
The observations were obtained in HST program GO9081 and are reported
elsewhere (Calvet et al. 2003b, in preparation). The low resolution (G140L) spectrum has been scaled to
100 AU, using 140 pc as the distance of Taurus \citep{kdh94},
and corrected for reddening 
using the interstellar law and $A_V$ = 0.5 (Gullbring et al. 1998).

We also show in Figure \ref{tts} the spectrum of the 10 Myr old
star TW Hya, from Herczeg et al. (2002). 
This star has a mass of 0.8 $\msun$, a radius of 1 $\rsun$ (Webb et al. 1999)
and a mass accretion rate $\sim 10^{-9} \msunyr - 10^{-8} \msunyr$
(Alencar \& Batalha 2002).
The flux from TW Hya suffers little or no reddening, but has been scaled 
by $\sim$ 3.5 to match the continuum of
BP Tau.\footnote{This is consistent with the ratio of their mass accretion 
rates, allowing for intrinsic variability and uncertainties of the
mass accretion rate determination.}
Comparison between the spectra of BP Tau and TW Hya indicates
that the shape of the continuum is roughly similar. More
differences can be seen in the strength of the emission 
lines which permeate the spectra. These are lines of highly ionized metals,
He, and Hydrogen Lyman series \citep[]{valenti,ardila}. The 
largest difference can be seen in
Lyman~$\alpha$, shown in the right inset in Figure \ref{tts}. 
The core of the line is lost to absorption in the line of sight 
to BP Tau. We also
show the Lyman~$\alpha$ profile of CY Tau,
a star with similar parameters to BP Tau
(Gullbring et al. 1998), but a lower reddening, $A_V$ = 0.3, so
more of the Lyman~$\alpha$ core can be seen. Still, the 
flux in Lyman~$\alpha$ is much lower than that of TW Hya.
This comparison indicates variations of more than a factor of 10 are expected
in the flux of the Lyman~$\alpha$ line among T Tauri stars.

With this information, we have constructed a composite
FUV  spectrum that is representative of the low mass 
T Tauri stars, which constitute the largest fraction
of the newly born young stars. The adopted spectrum is
equal to that of BP Tau between 1150 and 2000 {\AA}, 
and to the scaled TW Hya FUSE spectrum down to 950 {\AA}.
Below 950 {\AA}, we assume a linear extension with the same
slope as the continuum. For the chemical calculations in \S \ref{sec_chemical},
we integrate the adopted spectrum in 10 {\AA} bins. 

We show in Figure \ref{tts} the ISRF
scaled by 540 to match the total luminosity of the
adopted spectrum between 900 and 1700 {\AA}.\footnote{To make 
this comparison, we have
used the expression for the interstellar field energy density $u_\lambda$ from
Draine (1978) as given by Eq. (23) of Draine \& Bertoldi (1996),
with $J_\lambda = c u_\lambda / 4 \pi$.   In addition we have adopted 
the scaling $\chi = 1$,
which matches the interstellar field as estimated
by Habing (1968). This field is lower by a 
factor of 1.71 than the one given by $h c I_\lambda / 4 \pi \lambda$,
with $I_\lambda$ from van Dishoeck (1987). We have then
used $F_\lambda = 4 \pi J_\lambda = \int I_\lambda d \Omega$,
with $\Omega$ the solid angle, to compare with the
observed flux.}
The overall strength of the representative TTS FUV continuum spectrum at 100 AU
is of the order of a few hundred Habing, significantly lower than
generally assumed. In addition,
Figure \ref{tts} shows that the representative FUV spectra 
differs from the interstellar field in several
ways: (1) the shape of the continuum is different
in the sense that the TTS spectra rises as 
wavelength increases; (2) the TTS spectra
is dominated by emission lines, which carry at minimum $\sim$ 35 \% of
the flux.\footnote{
For BP Tau and TW Hya lines carry, respectively, at least 35\% and 85\%
of the luminosity between 1100--1700 \AA.  These percentages are lower limits
because of interstellar and wind absorption of Lyman $\alpha$
photons.}

\section{Disk Model}
\label{sec_model}

For the calculations in this paper,
we adopt a disk model which fits the median
spectral energy distribution
(SED) for classical (accreting) T Tauri stars (TTS) in the Taurus clouds
\citep{dch01}.
Typical stellar parameters are adopted,
$M_*=0.5 \ \msun$, $R_*=2 \ \rsun$
and $T_*=4000$ K \citep{gull98,HCGD}, and the mass
accretion rate used, $\mdot=3 \times 10^{-8} \msunyr$, is near the
average in Taurus \citep{HCGD}. Dust grains
were assumed to be segregate spheres of compounds with
abundances as in Pollack et al. (1994), and with a size distribution
given by $n(a) da \propto a^{-3.5} da$, where the grain radius $a$
varies between $0.005 \ \mu$m and 1 mm.
The disk outer radius is $R_d=$ 350 AU; with this,
and $\alpha = $ 0.05,
the mass of the disk is
0.03 $\msun$.

\begin{figure}
\plotone{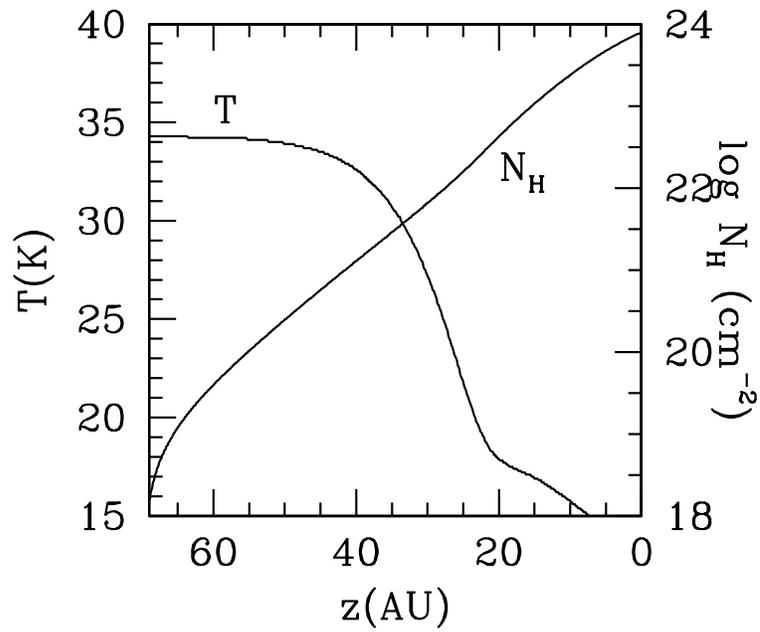}
\caption{Temperature and column density as a function of
height $z$ at disk radius $R $ = 100 AU for the disk
model used in this work, which fits the
median SED of Taurus stars (see \S \ref{sec_model}).
}
\label{estructura}
\end{figure}

We self-consistently solve the complete set of vertical structure
equations, including irradiation and viscous heating, resulting
in detailed profiles of temperature and density with
vertical height and disk radius. 
Figure \ref{estructura} shows the vertical profiles of
temperature and column density for the adopted disk model at a radius
$R$ = 100 AU, which will be used in the discussions of
the following sections.

\section{Transfer of FUV radiation}
\label{sec_uvtransf}

\begin{figure}
\plotone{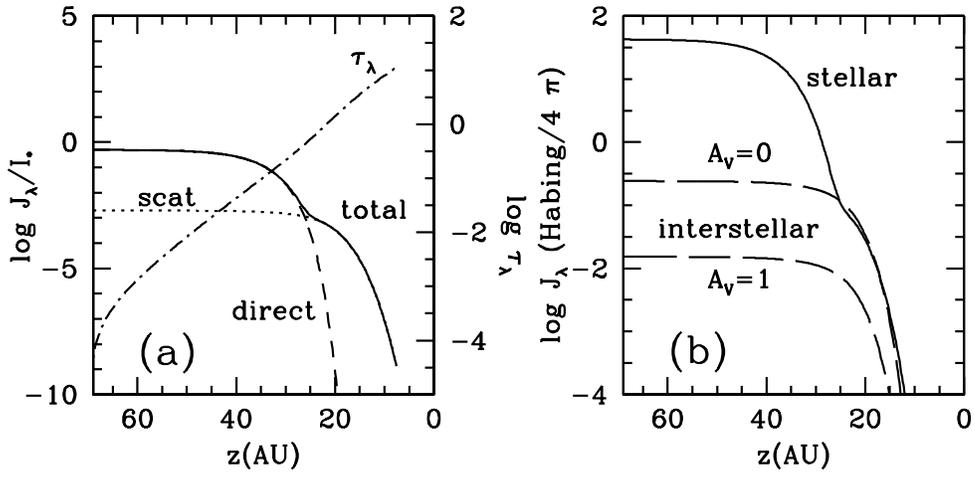}
\caption{(a) Normalized mean intensity at $\lambda$ = 1500 {\AA} 
and disk radius 100 AU
as a function of disk height z.
The total mean intensity
(solid line) is separated into the direct stellar component (dashed line)
and the diffuse scattered component (dotted line). The vertical
optical depth is also shown for reference (dot-dashed line). The mean intensities
are normalized to the incoming stellar specific intensity
at the same wavelength.
For the dust mixture considered in this work,
the scattering coefficient 
$\sigma_{\lambda}$ goes from $\sim$ 0.4 at 1000 - 1400 {\AA} to
$\sim$ 0.8 at 2000 - 3000 {\AA}.
(b) Comparison between the mean intensities penetrating
the disk at $\lambda$ = 1500 {\AA} and $R = $ 100 AU:
stellar field, with an incoming specific intensity I$_*$ corresponding to the
continuum at 1500~\AA\ in Figure~1; 
interstellar field (dashed lines), assuming 
A$_V$ = 0 and 1 mag.
}
\label{jotas}
\end{figure}

We follow the transport of FUV radiation through the disk
using an analytical approximation that allows for the 
inclusion of scattering in addition to pure absorption 
\citep[see also][]{vanz_zchem}.
We assume that
stellar radiation at an ultraviolet wavelength $\lambda$ comes in a single beam which makes
an angle $\theta_0(\lambda)$ to the local surface $z_s(\lambda)$,
defined by the condition $\tau_{\lambda}^{rad} \sim 1$, where
$\tau_{\lambda}^{rad}$ is the radial optical depth. 
We follow standard procedures for solving the transfer
equations in the plane-parallel approximation\footnote{
This is justified because of the small geometrical
thickness of the disk (cf. D'Alessio et al. 1999).}. We assume that there is
no local emission at $\lambda$. Along the original beam, defined by 
$\mu_0 = \cos (\theta_0)$\footnote{At R = 100 AU $\mu_0(1000$\AA) = 0.07.},
the mean intensity decreases as $exp (- \tau_{\lambda} / \mu_0)$,
where $\tau_{\lambda}$ is the vertical optical depth. A nearly
isotropic diffuse field is created by photons scattered out of the beam.
Following \cite{M78}, we solve the transfer equations for
symmetric and antisymmetric  averages
of the specific intensity $u$ and $v$, from which we can derive 
with usual boundary conditions the depth
dependence of the mean intensity as
\begin{equation}  
J_\lambda =  
{I_* \over 2}  C_1  e^{- \tau_{\lambda}  / \mu_0}  +
\sigma_{\lambda} I_* \mu_0  C_2 e^ { \sqrt{3 ( 1 - \sigma_{\lambda})} \tau_{\lambda}}
\label{jnu}
\end{equation}
where
\begin{equation}
\begin {array}{c}
C_1 = 1 - 3 \sigma_{\lambda} \mu_0^2 \\
C_2 = { { 1  + 3/2 \, \mu_0 + 3 ( 1 - \sigma_{\lambda}) \mu_0^2} \over
{ 1 + 2 \sqrt{(1 - \sigma_{\lambda})/3} } }.
\end{array}
\end{equation}
and $\sigma_{\lambda}$ is the albedo.
The first term in Eq. (\ref{jnu}) corresponds to
the direct incoming beam that is rapidly attenuated due to the
oblique angle of incidence. 
The second term is the diffuse scattered radiation field which 
penetrates deeper than the direct radiation.
The mean intensity $J_\lambda$
at $\lambda$ = 1500 {\AA}, normalized to the 
stellar intensity $I_*$ at the same
wavelength, is shown in Figure \ref{jotas}a, where each
contribution 
is shown separately. 
Similarly to the models of \citet{vanz_zchem}, the scattered field penetrates 
much closer to the midplane than the direct attenuated stellar component.
This holds true to within a few hundred AU of the star. 

The stellar specific intensity in Eq.~(\ref{jnu}) can be estimated
assuming that the beam carries a flux equal to the
observed flux $F_\lambda$, so
$I_* = F_\lambda/ 2 \pi$. Figure \ref{jotas}b
shows the mean intensity $J_\lambda$ at $\lambda$ = 1500~{\AA}
and $R$ = 100 AU from Figure \ref{jotas}a,
scaled by $I_*$ calculated 
from the continuum flux at 1500 {\AA}
in Figure \ref{tts}.
We also plot for comparison
the mean intensity corresponding to the ISRF (1 Habing) impinging
on the disk surface with $\mu_0 = 1$ in Eq.(\ref{jnu}). 
It can be
seen that the stellar field dominates in the upper $\sim$ 30 AU
in the disk, while it becomes comparable or smaller than
the interstellar field 
closer to the midplane \citep{willacy_zchem, vanz_zchem}.  However,
TTS in Taurus
have a mean extinction $A_V \sim 1 $ along the line
of sight \citep{KH95}, which corresponds to an attenuation $\sim$ 16 
in the UV; the corresponding mean
intensity is shown in Figure \ref{jotas}b.
Moreover, radiation in the lines can be much higher than
in the continuum (cf. Figure \ref{tts}).
Therefore, the stellar field
is expected to be the dominant contributor
to the FUV field illuminating the disk \citep[see][]{aikawa_zchemav}.

\section{Lyman $\alpha$ and Molecular Photodissociation}
\label{sec_chemical}

Using the observed radiation field in Figure~\ref{tts} we can
examine the effects of Lyman~$\alpha$ on the photodissociation 
of CN, HCN, and H$_2$O.   To calculate the photodissociation rate we use
the following expression 

\begin{equation}
k_{pd}(s^{-1}) =  \int
{4\pi \lambda \over hc}  \sigma (\lambda )J_{\lambda } d\lambda.
\end{equation}

For the radiative transfer we use
the procedure of \S \ref{sec_uvtransf} and the disk model
in \S \ref{sec_model}.
In this calculation the Lyman~$\alpha$
flux is initially assumed to be a constant value between the peaks of the 
red and blue damping wings of the CY Tau profile (Figure~\ref{tts} 
right, \S \ref{sec_fuv}).  As this is a lower
limit, in one case we increase the flux of the Lyman~$\alpha$ 
feature between 1205 - 1230 \AA\  
by a factor of 10.   This increases the total UV flux by a factor of three.
For comparison we provide the corresponding photodissociation
rate from a scaled interstellar FUV field that is attenuated only (i.e. without 
a scattered contribution, similar to Willacy \& Langer 2000). 

\begin{figure}
\epsscale{1.0}
\plotone{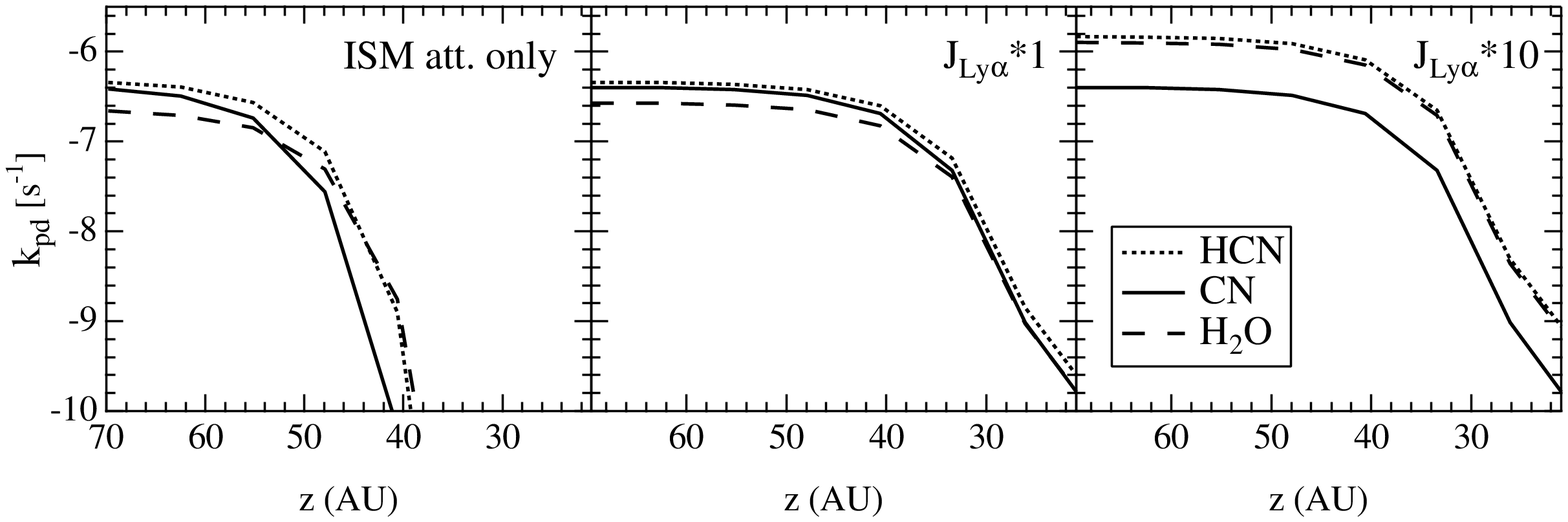}
\caption{
Photodissociation rate for CN, HCN, and H$_2$O 
calculated using FUV photoabsorption crossections 
from \citet{nuth_hcnxs}, \citet{lavendy_cnxs},
\citet{nee_cnxs}, \citet{yoshino_h2oxs}, and \citet{chan_h2oxs}.
Three models are
presented. (1) ISM att. only: using the ISM photodissociation rate taken from \citet{millar_net}
scaled by $\sim 350$. This rate only includes attenuation within the disk model
in \S 3 and 
no contribution from the diffuse scattered field.
(2) J$_{Ly \alpha}$*1 using FUV radiative transfer 
as in \S \ref{sec_uvtransf}, disk model from \S 3, and FUV field from Figure~1.
The Lyman~$\alpha$ flux  is assumed to be a constant value between 
the red and blue damped wings shown
in Figure~\ref{tts}.
(3) J$_{Ly \alpha}$*10: same as (2) but
the flux between 1205 - 1230 \AA\ is increased 
by a factor of 10.  Because there are no known CN photodissociation channel above
1150 \AA\ this increase results in no change in the CN photodissociation rate, while 
the HCN and H$_2$O rates are drastically affected. 
}
\label{photorate}
\end{figure}

Our results are given in Figure~\ref{photorate}.
Two main factors are noticed in this plot. (1) For all species 
the inclusion of scattering allows for greater FUV penetration than in models which simply attenuate
a parameterized ISRF (van Zadelhoff et al. 2002).
(2) For  HCN and H$_2$O the presence of an enhanced Lyman~$\alpha$ radiation field
can increase photodissociation. 
Because of the difference in photodissociation channels, 
CN will exist over a larger range of vertical distances within the disk than either HCN or H$_2$O. 

\section{Discussion}
\label{sec_discussion}

While the primary photoabsorption bands for CO and H$_2$ lie short of 1100
\AA\ numerous other molecules will be subject to photodissociation by 
Lyman~$\alpha$ radiation \citep[e.g. OH, CH$_4$, NH$_3$;][]{lepp_xray}.  Thus, 
the presence of  Lyman~$\alpha$ photons can impact the chemistry
on the disk surface beyond the CN/HCN ratio.   Indeed FUV radiation will 
be of increasing importance in disk planet forming regions and Lyman~$\alpha$
emission may be an important factor in limiting the abundance of water vapor. 
Young stars are also
show evidence for significant emission from X-rays, which can ionize and
dissociate H$_2$ molecules.  This process results in the creation of 
an {\em in situ} Lyman~$\alpha$
radiation field that can further alter the chemistry \citep{lepp_xray}.\footnote{
Several effects can moderate the importance of Lyman $\alpha$ radiation.
The presence of a large column of hydrogen atoms will predominantly scatter Lyman $\alpha$
photons.   This will result in the destruction of some photons through
dust absorption.  Using the calculations of Bonilha et al. (1979) we estimate
that N$_H$ $\sim 10^{21}$ cm$^{-2}$ is required to significantly attenuate
the emission. 
In the inner disk, a water column of $\sim 10^{18}$ cm$^{-2}$ will be
sufficient to extinguish the Lyman $\alpha$ line core radiation;
larger columns are needed to absorb 
both line core and wing emission. 
This will not happen in the outer disk, where water is frozen on grains.
Finally, H$_2$ atoms can also absorb Lyman $\alpha$ photons resulting in
fluorescence emission.   
}

Furthermore the FUV radiation field
representative of low mass T Tauri stars with an average mass accretion
rate 
is weaker than generally assumed in chemical calculations.  
The origin of the excess FUV flux is still under
investigation, but in the study of a sample of HST/STIS spectra, the
luminosity in the FUV is found to scale with the accretion luminosity
(Calvet et al. 2003, in preparation).
Since the mass accretion rates of the objects with the
majority of molecular detections,
DM Tau and LkCa15, 
are comparable and below that of 
BP Tau, respectively \citep{HCGD},
this suggests that the FUV radiation field in these objects may be 
similar or weaker than BP Tau.   With the stronger FUV field
current models are capable of reproducing observed abundances of species such as CN and
HCN in DM Tau (Aikawa et al. 2002); assuming a weaker FUV field may make it difficult
to continue to match observations, but strong  
Lyman~$\alpha$ radiation can provide an additional source of photodissociation to
power the chemistry. 

\acknowledgements

We are grateful to an anonymous referee for useful comments.
This work has been supported in part by
NASA through grants GO-08206.01-A, 
GO-9081.01-A, and GO-09374.01-A from the Space Telescope
Science Institute and Origins of Solar Systems grant NAG5-9670.
PD acknowledges grants from DGAPA and CONACyT.


\end{document}